\documentclass{elsart}

\usepackage{cite}
\usepackage{epsfig}
\usepackage{url}
\usepackage{amssymb}
\def\dd{.\kern-1pt.\kern-1pt.}
\def\pic{\ifm{\pi^+\pi^-}} 
\def\ab{\ifm{\sim}} 
\def\ppg{\pi^{+}\pi^{-}\gamma}
\def\ppp{\pi^{+}\pi^{-}\pi^0}
\def\eeg{e^{+}e^{-}\gamma}
\def\mmg{\mu^{+}\mu^{-}\gamma}
\def\ifm#1{\relax\ifmmode#1\else$#1$\fi}
\def\epm{\ifm{e^+e^-}} 
\def\to{\ifm{\rightarrow}}
\def\dif{\hbox{d}} \def\sig{\ifm{\sigma}} 
\def\gam{\ifm{\gamma}}  
\def\mtrk{m_{\rm trk}} \def\deg{\ifm{^\circ}}
\def\plm{\ifm{\pm}}  \def\km{\kern-1.5mm}  \def\kak{\km&\km}
 \def\epm{\ifm{e^+e^-}}

\newcommand{\AmS}{{\protect\the\textfont2
  A\kern-.1667em\lower.5ex\hbox{M}\kern-.125emS}}
\newcommand{\amu}{$a_\mu$}
\makeatletter
\newdimen\z@ \z@=0pt 
\newskip\z@skip \z@skip=0pt plus0pt minus0pt
\def\m@th{\mathsurround=\z@}
\def\ialign{\everycr{}\tabskip\z@skip\halign} 
\def\eqalign#1{\null\,\vcenter{\openup\jot\m@th
  \ialign{\strut\hfil$\displaystyle{##}$&$\displaystyle{{}##}$\hfil
      \crcr#1\crcr}}\,}
\makeatother
\newcommand{\aff}[2]{Dipartimento di Fisica dell'Universit\`a #1 e 
Sezione INFN, #2, Italy.}
\newcommand{\affd}[1]{Dipartimento di Fisica dell'Universit\`a e 
Sezione INFN, #1, Italy.}

\begin{document}

\begin{frontmatter}

\title{Measurement of $\sigma(e^+e^-\rightarrow \pi^+ \pi^- \gamma)$ 
and extraction of  $\sigma(e^+e^-\rightarrow \pi^+\pi^-)$ 
 below 1 {\rm GeV} with the KLOE detector}

\vskip -1cm 

\collab{The KLOE Collaboration}

\author[Na] {A.~Aloisio},
\author[Na]{F.~Ambrosino},
\author[Frascati]{A.~Antonelli},
\author[Frascati]{M.~Antonelli},
\author[Roma3]{C.~Bacci},
\author[Roma3]{M.~Barva},
\author[Frascati]{G.~Bencivenni},
\author[Frascati]{S.~Bertolucci},
\author[Roma1]{C.~Bini},
\author[Frascati]{C.~Bloise}
% ...

\clearpage
\address[Bari]{\affd{Bari}}
\address[Beijing]{Permanent address: Institute of High Energy 
Physics, CAS,  Beijing, China.}
\address[Frascati]{Laboratori Nazionali di Frascati dell'INFN, 
Frascati, Italy.}
\address[Karlsruhe]{Institut f\"ur Experimentelle Kernphysik, 
Universit\"at Karlsruhe, Germany.}
\address[Lecce]{\affd{Lecce}}
\address[Moscow]{Permanent address: Institute for Theoretical 
and Experimental Physics, Moscow, Russia.}
\address[Na]{Dipartimento di Scienze Fisiche dell'Universit\`a 
``Federico II'' e Sezione INFN,
Napoli, Italy.}
\address[Budker]{Permanent address: Budker Institute of Nuclear Physics, 
Novosibirsk, Russia.}
\address[Pisa]{\affd{Pisa}}
\address[Energ]{Dipartimento di Energetica dell'Universit\`a 
``La Sapienza'', Roma, Italy.}
\address[Roma1]{\aff{``La Sapienza''}{Roma}}
\address[Roma2]{\aff{``Tor Vergata''}{Roma}}
\address[Roma3]{\aff{``Roma Tre''}{Roma}}
\address[StonyBrook]{Physics Department, State University of New 
York at Stony Brook, USA.}
\address[Virginia]{Physics Department, University of Virginia, USA.}

\begin{abstract}
We have measured the cross section 
$\sigma(e^+e^-\rightarrow \pi^+\pi^- \gamma)$ 
at an energy $W=m_\phi=1.02$ GeV with the KLOE
 detector at the electron-positron 
collider DA$\Phi$NE. From the 
dependence of the cross section on the invariant mass of the two-pion system,
we extract $\sigma(e^+e^-\rightarrow \pi^+\pi^- )$ for the mass range 
$0.35<s<0.95$ GeV$^2$. From this result, we
calculate the pion form factor and
the hadronic contribution to the muon anomaly, $a_\mu$.
\end{abstract}

\end{frontmatter}

\section{Hadronic cross section at DA$\Phi$NE}

\subsection{Motivation}

The recent precision measurement of the muon anomaly \amu\ 
at the Brookhaven National Laboratory~\cite{Bennett:2004pv}
has led to renewed interest in an accurate measurement of the cross
section for $e^+e^-$ annihilation into hadrons. Contributions 
to the photon spectral functions due to quark loops 
are not calculable for low-hadronic-mass states by perturbative QCD at low energy. 
However, they can be obtained by connecting  
the 
imaginary part of the hadronic piece of the polarization function 
by unitarity to the cross section for $e^+e^-
\rightarrow$ hadrons~\cite{durand,gourdin}. A dispersion relation can thus be derived,
giving the contribution to \amu\ as an integral over the hadronic 
cross section multiplied by a kernel $K(s)$, which behaves approximately like $1/s$: 
\begin{equation}
a_\mu^{\rm had}={1\over4\pi^3}\int_{4 m_{\pi}^2}^{\infty}%
\sigma_{\epm\to{\rm hadr}}(s)K(s)\dif s.
\label{eq:amu}
\end{equation}
The process \epm\to\pic\ below $1$ GeV accounts for  $62\%$ of the 
total hadronic contribution~\cite{Eidelman:1995ny}.
The most recent measurements of
\sig(\epm\to\pic) for values of $\sqrt{s}$ between 
610 and 961 MeV come from the CMD-2 experiment at 
VEPP-2M where the quoted systematic error is 0.6\% and the contribution of
the statistical error on $a_\mu^{\rm had}$ is \ab0.7\% \cite{Akhmetshin:2001ig,Akhmetshin:2003zn}.
These data, together with $\tau$ and \epm\ 
data up to 3 GeV,
 have been used to produce a prediction for 
comparison with the BNL result~\cite{Davier:2003pw}. 
There is however a rather strong disagreement 
between the $a_\mu^{\rm had}$ value obtained using $\tau$ decay data 
after isospin-breaking corrections and
\epm\to\pic\ data. 
Moreover, the \epm\to\pic\ based result disagrees by \ab3\sig\ 
with the direct 
measurement of \amu . 

\subsection{Radiative Return}

Initial state radiation (ISR) is a convenient mechanism 
that allows one to study \epm\to\ hadrons over
 the entire range from 2$m_\pi$ to $W$, the center of mass 
energy of the colliding beams. 
In this case, there are complications from  final-state 
radiation (FSR). For a photon radiated 
prior to the annihilation of the \epm\ pair, the mass of the \pic system  
is\setcounter{footnote}{0}\footnote{Neglecting the small momentum of the 
$\phi$.}
 $m_{\pi^+\pi^-}=\sqrt{W^2-2WE_\gam}$. Instead, for a photon radiated 
by the final-state pions, the virtual photon coupling to the \pic\ 
pair has a mass $W$. By counting vertices, the relative probabilities of ISR and FSR are of the same order. 
This requires careful estimates of the 
two processes in order to be able to use the reaction \epm\to\pic\gam\ 
to extract \sig(\epm\to\pic). The Karlsruhe theory group has developed the EVA and PHOKHARA Monte Carlo 
programs which are fundamental to our analysis 
\cite{binner,german,czyz,Kuhn:2002xg,grzelinska,fsrczyz}. 
In particular, the PHOKHARA Monte Carlo simulation has been used to evaluate
the contribution for the ISR process (via the radiation function $H$) 
in order to derive the hadronic cross section:

\begin{equation}
s_\pi \frac{d\sigma_{\pic\gamma}}{ds_\pi} =
\sigma_{\pic}(s_\pi)  H(s_\pi),
\label{eq:H}
\end{equation}
where $s_\pi =m^2_{\pi^+\pi^-}$,
which coincides with the invariant mass $s$ of the intermediate
photon for the case of ISR radiation only. 
The equation above is correct at leading order if FSR emission
can be neglected. The case of NLO terms, with the simultaneous
emission of ISR and FSR photons, is discussed in Sec.~\ref{FSR}.

The present analysis
is based on the observation  of 
Ref.~\cite{binner} that for small polar angle $\theta_\gamma$ of the radiated photon, 
the ISR process vastly dominates over the FSR process. 
In the following we restrict ourself to studying the 
reaction \epm\to\pic\gam\ with $\theta_\gamma<15^\circ$ or 
$\theta_\gamma>165^\circ$. For small $s_{\pi}$, the di-pion 
system recoiling against a small angle photon will result in 
one or both pions being lost at small angle as well. 
We are therefore  limited to measuring \sig(\pic) for $\sqrt{s_\pi}>$550 MeV. 
In the future extension of this work
we will be able to measure the cross section near threshold. 
This is very important, since there are no 
good measurements of \sig(\pic) at low masses, 
which weigh strongly in the estimate of 
$a_\mu^{\rm had}$.

\section{Measurement of $\sigma(e^{+}e^{-} \rightarrow \ppg)$}
\label{}

\subsection{Signal selection}
\begin{figure}
\begin{center}
\mbox{\epsfig{file=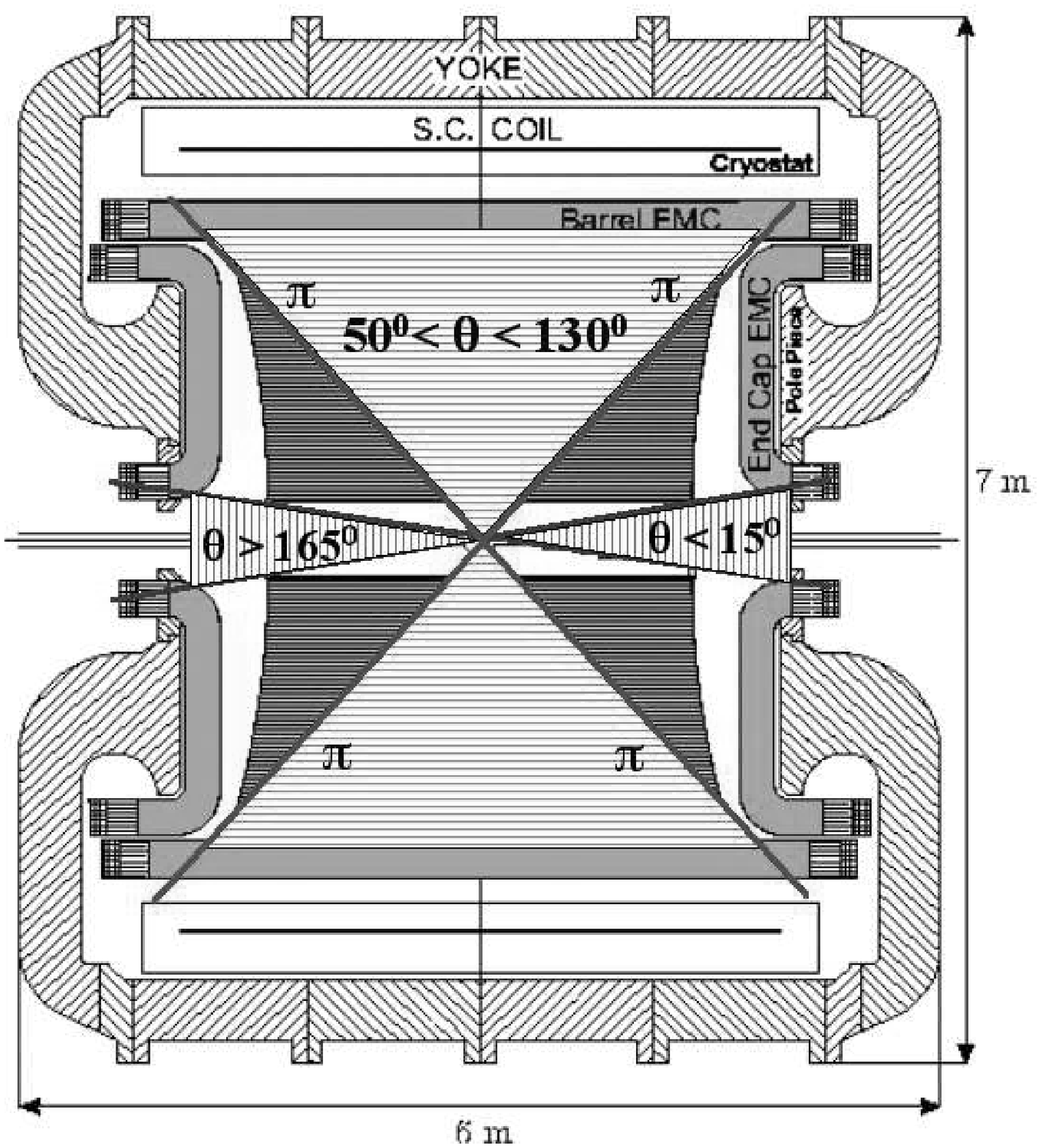,width=6cm}}
\mbox{\epsfig{file=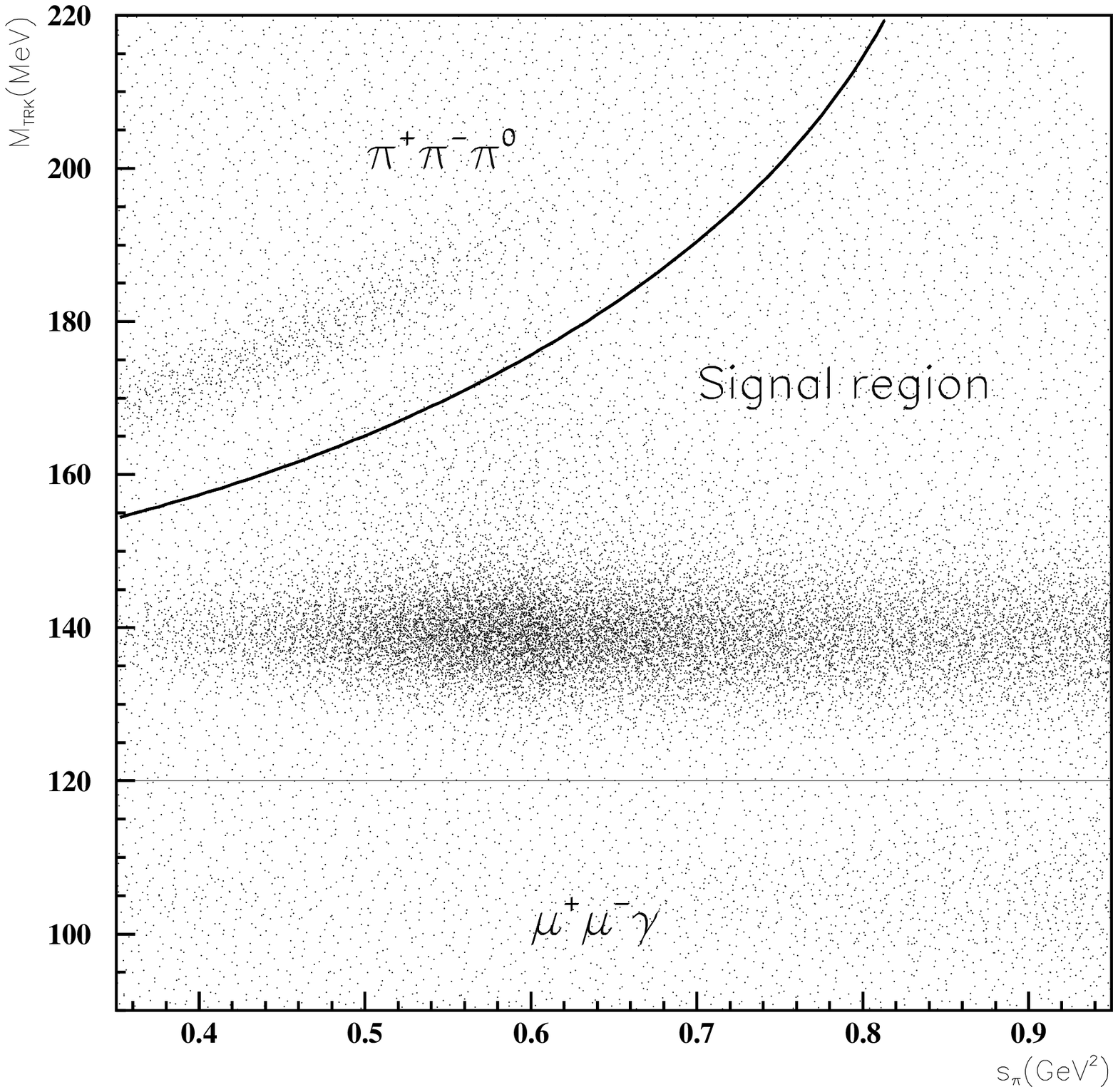,width=7cm}}
\caption{Left: schematic view of the KLOE detector with the angular acceptance regions for pions (\textit{horizontally hatched area}) and photons (\textit{vertically hatched area}). The photon angle is evaluated from the two pion tracks.
Right: 2-dimensional requirement in the plane of $m_{\rm trk}$/MeV and $s_\pi$/$GeV^2$.}
\label{fiducial}
\end{center}
\end{figure}
The KLOE detector consists mainly of a high resolution drift chamber with transverse momentum resolution 
$\sigma_{p_T} / p_T \leq 0.4\%$~\cite{dc} and an electromagnetic calorimeter with energy resolution 
$\sigma_E / E = 5.7 \% /  \sqrt{E({\rm GeV})}$~\cite{emc}. 
In the current analysis, we have concentrated on events in which the pions are emitted at polar angles $\theta_\pi$ between $50^\circ$ and $130^\circ$. 
The direction and energy of the photon is reconstructed from the pion tracks by closing the kinematics; explicit photon detection is not required.
As a consequence, a requirement on the di-pion production angle $\theta_{\pi\pi}$ smaller than $15^\circ$ (or greater than $165^\circ$) is performed instead of a requirement on the photon angle $\theta_\gamma$. 
The acceptance regions are shown in Fig.~\ref{fiducial}, left. 
These specific acceptance requirements reduce background contamination and the
relative contribution of final-state radiation from the pions
to very low levels~\cite{Cataldi}. It will be shown in the following that an efficient and nearly background free signal selection can be done without explicit photon tagging. 

The selection of $e^{+}e^{-} \rightarrow \ppg$ events
is performed with the following steps:
\begin{itemize}
\item \textit {Detection of two charged tracks connected to a vertex}: Two charged tracks with polar angles between $50^\circ$ and $130^\circ$ connected to a vertex in the fiducial volume, $R_{xy}<8$ cm, $|z|< 7$ cm, are required. Additional requirements 
on transverse momentum, $p_{T} > 160$ MeV, and on longitudinal momentum,
 $|p_{z}| > 90$ MeV, reject spiralling tracks and ensure good reconstruction conditions.
\item \textit {Identification of pion tracks}: Separation of pions from electrons is performed  using a PID method based on approximate likelihood estimators. 
These estimators are based on the comparison of
time-of-flight versus momentum and on
the shape and energy deposition of the 
calorimeter clusters produced by charged tracks. The functions have been built using control samples of  $\pi^{+}\pi^{-}\pi^{0}$ and $\eeg$ events in data, in order to obtain the calorimeter response for pions and electrons. 
An  event is selected as signal if at least one of the two tracks is identified as a pion. In this way, the content of $\eeg$ events  in the signal sample 
is drastically reduced,
while the efficiency for retaining $\ppg$ events is still very high ($>$ 98\%).
 \item \textit {Requirement on the track mass}: The track mass ($m_{\rm trk}$) is a kinematic variable corresponding to the mass of the charged tracks  under the hypothesis that the final state consists of two particles with the same mass and one photon. 
It is calculated from the reconstructed momenta of the $\pi^+$ and $\pi^-$ ($\vec{p}_{+}$, $\vec{p}_{-}$) and the 
center-of-mass energy $W$. 
Requiring a value larger than 120 MeV rejects $\mmg$ events, while in order to reject  $\pi^{+}\pi^{-}\pi^{0}$ events, an $s_\pi$-dependent requirement is used
 (see Fig.~\ref{fiducial}, right).  
 \item \textit {Requirement on the di-pion angle $\theta_{\pi\pi}$}: The aforementioned requirement on the di-pion angle $\theta_{\pi\pi} < 15^\circ$ or $> 165^\circ$ 
and $50^\circ<\theta_{\pi}<130^\circ$ is performed. 
\end{itemize}
\begin{figure}[htp!]
\begin{center}
\mbox{\epsfig{file=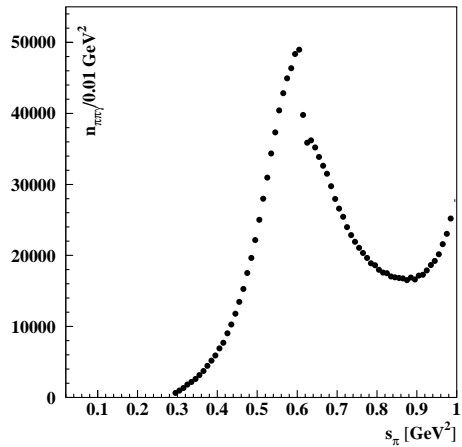,width=6.5cm}}
\caption{Distribution of counts as a function of $s_\pi$,
in bins of 0.01 GeV$^2$,
after applying the acceptance and selection requirements. ${\mathcal L}=141.4~\mathrm{pb}^{-1}$;
data from 2001.}
\label{fig:nraw}
\end{center}
\end{figure}
The data used for the analysis were taken from
 July to December 2001, yielding an integrated luminosity of
 ${\mathcal L}=141.4~\mathrm{pb}^{-1}$. After the selection requirements  
mentioned above,
 we find $~1.555 \times 10^6$ events, corresponding to $\simeq
 11000~\mathrm{events/pb}^{-1}$. Figure~\ref{fig:nraw}  shows the
 distribution of the $\ppg$ events in bins of 0.01 GeV$^2$ for
 $s_\pi$.  The $\rho$ peak and the $\rho - \omega$ interference
 structure are clearly visible, even without unfolding the spectrum
 from the detector resolution, demonstrating the excellent momentum
 resolution of the KLOE detector.  To obtain the cross section for
 $0^\circ<\theta_\pi<180^\circ$ and $\theta_{\pi\pi}<15^\circ$, 
$\theta_{\pi\pi}>165^\circ$ we
 subtract the residual background from this spectrum and divide by the
 selection efficiency, acceptance, and integrated
 luminosity:

\begin{equation}
\frac{d\sigma_{\ppg}}{ds_\pi}=
\frac{\Delta N_{\rm Obs}-\Delta N_{\rm Bkg}}{\Delta s_\pi}  
\frac{1}{\epsilon_{\rm Sel}\epsilon_{\rm Acc}} 
\frac{1}{\int{\mathcal{L}} dt}.
\label{eq:sighad}
\end{equation}

The background subtraction, the evaluation of the selection efficiency and the acceptance, the measurement of the integrated luminosity, and the unfolding of the experimental resolution on $s_\pi$ (omitted from Eq.~(\ref{eq:sighad}) for clarity)
are discussed below. Detailed information on all the aspects of the analysis is available in~\cite{kn}.
\subsection{Background subtraction}

\begin{figure}[ht]
\begin{center}
\mbox{\epsfig{file=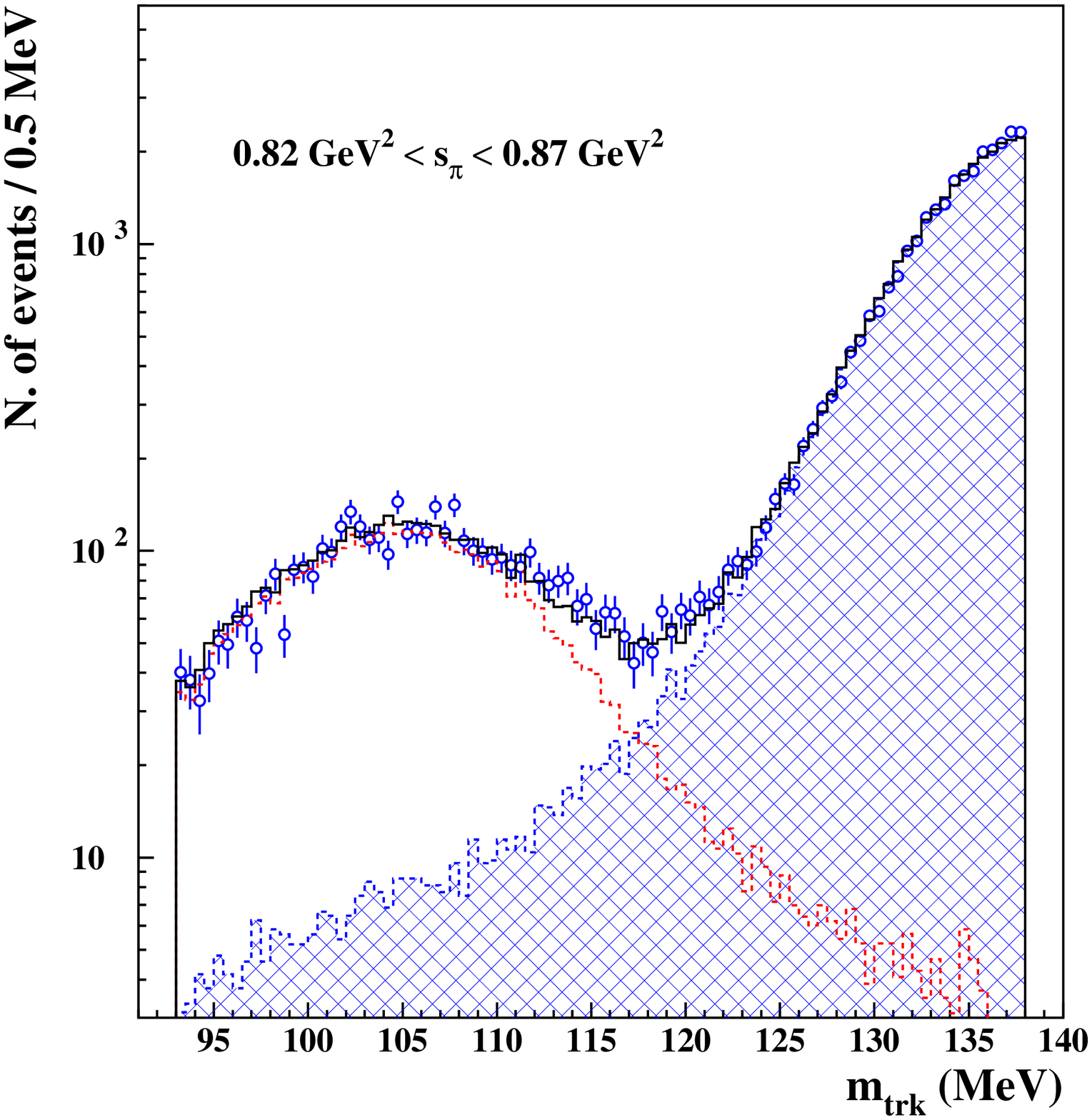,width=6.5cm}}
\mbox{\epsfig{file=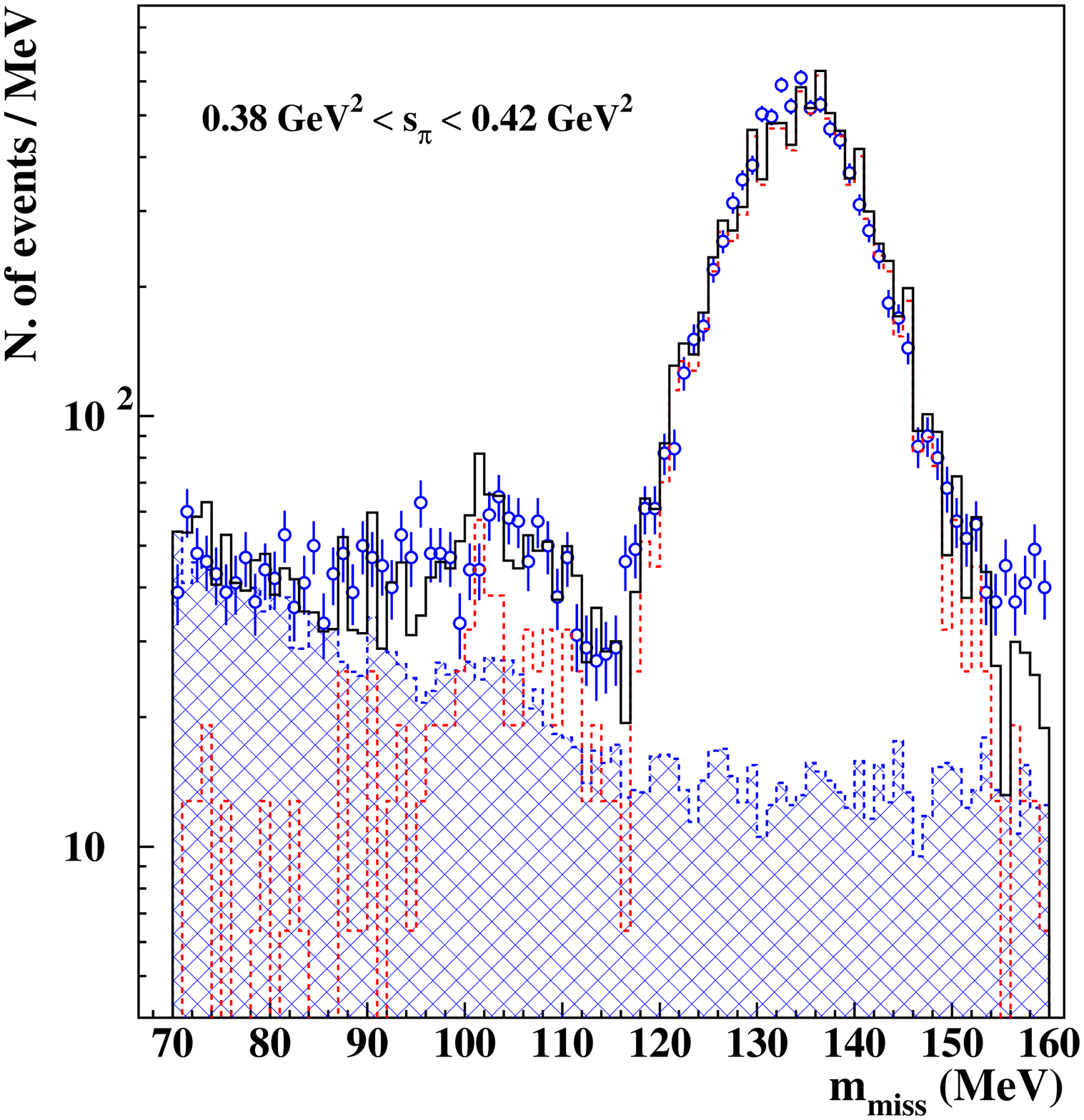,width=6.5cm}}
\caption{Left: fit of the mass peak for muons. Right: $\pi^0$ mass fit.
These fits are used to estimate the $\mmg$ and $\ppp$ backgrounds to 
 the $\ppg$ channel. Points are data, solid line is Monte Carlo simulation.
Dashed line and hashed area represent the
 Monte Carlo evaluation of background ($\mmg$ in the left plot and $\ppp$ in the right one) and $\ppg$ contributions, respectively. }
\label{fig:bkg1}
\end{center}
\end{figure}

After applying the requirements on the fiducial volume, the likelihood, and
$m_{\rm trk}$, a residual background of $\eeg$ , $\mmg$, and $\ppp$ events remains.
The  population of  signal 
and background events in the [$s_\pi$,$\mtrk$] plane is illustrated in Fig.~\ref{fiducial}, right.
Background from $\eeg$ and $\mmg$ events is concentrated at
low values of $\mtrk$ . 
The amount of background 
in the signal region is obtained by fitting the $\mtrk$
spectra of the selected events (except for the $m_{\rm trk}$ requirement) 
in slices of $s_\pi$.
The $\mtrk$ spectra for signal and  $\mmg$
events are obtained from  Monte Carlo simulation, 
while for $\eeg$  events, $\mtrk$ is obtained directly from data,
using a dedicated sample of 152 pb$^{-1}$. 
 An example of such a fit 
to determine the  background fraction for  $\mmg$ events
is shown in Fig.~\ref{fig:bkg1}, left.
Background from $\ppp$ events appears at higher 
$\mtrk$ values and the missing mass 
$m_{\rm miss}^2 = ({p_\phi} - {p}_+ - {p}_-)^2$, 
peaks at $m_{\pi^0}^2$. The number of $\ppp$ events in
the signal region is obtained by fitting the $m_{\rm miss}$
distribution with the shapes obtained from the Monte Carlo
simulation; an example is shown in Fig.~\ref{fig:bkg1}, right.
%
The shape of the background distribution is well
reproduced by the Monte Carlo simulation, ensuring that systematic
uncertainties are smaller than the fit errors, which are
considered as systematic errors of the procedure. 

The contribution of all backgrounds to the observed
 signal is  below 2\% above 0.5~GeV$^2$, while it increases up to $\sim$ 10\%
at $s_\pi = 0.35$  GeV$^2$.
%
%
Other possible sources of background for which the contributions have been evaluated are the process
$e^+e^- \to e^+e^-\pi^+\pi^-$ with the electrons
emitted along the beam pipe~\cite{nowak} and the NLO
corrections to the FSR in the process
$e^+e^- \to \mu^+\mu^-\gamma$~\cite{phokhara4}. 
The systematic uncertainties associated with the background estimates for
all these sources have been added
in quadrature; the results are shown in Table~\ref{tab:syseff0}.

\subsection{The selection efficiency $\epsilon_{\rm Sel}$}
 The selection efficiency is the product of the efficiencies associated
with the trigger, the event reconstruction, the background filtering 
and the track mass requirement:
 in 
\begin{itemize}
\item\textit  {Trigger efficiency}: 
Events in the $\ppg$ sample must satisfy the
calorimeter trigger, {\it i.e.}, there must be at least two
trigger sectors with energy deposition above threshold 
 (for details on the KLOE trigger
see~\cite{kloetrig}). 
The trigger also includes a cosmic-ray veto: events with energy deposition above a certain threshold in the outermost layer of the calorimeter are rejected online. 
While the calorimeter trigger itself  is rather efficient
for signal events ($>95\%$), the cosmic-ray veto rejects a significant 
fraction of
$\ppg$ events since such events mimic cosmic rays.
The cosmic-ray veto inefficiency is on the level of few percent at small
values of $s_\pi < 0.4 {\rm GeV}^2$ but reaches up to $30\%$ at 
$s_\pi = 0.95 {\rm GeV}^2$.
The overall
trigger efficiency, including the effect of the cosmic-ray veto, was evaluated 
from the probability for single pions to fire trigger sectors in 
$\ppg$ events wherein part of the event could be ascertained to have satisfied the trigger alone.
 The fractional uncertainty associated with this procedure was estimated to be 
$\delta\epsilon_{\rm TRG}(s_\pi) = \mbox[{\rm exp}(0.43-4.9s_\pi[{\rm GeV}^2]) + 0.08]$ (expressed in percent), and is 
dominated by the systematics of establishing the 
correct track-to-trigger sector association.
 
\item \textit  {Background filter efficiency}: 
During reconstruction, an of{}f{}line filtering procedure identifies 
and rejects background events as soon as 
 they have been reconstructed in the calorimeter~\cite{offline}. The efficiency of this filter has been studied using a dedicated sample of
$\ppg$ events that were rejected by the filter itself.
Since the filtering procedure is
 very sensitive to the presence of 
accidental clusters in the electromagnetic calorimeter, 
the  efficiency of this filter was parametrized as a function of
the background conditions during data taking and averaged over time.
The filter efficiency was found to be uniform in $s_{\pi}$ and 96.6\% on average,
with a flat systematic error of 0.6\%.

\item \textit {Tracking efficiency}: 
The tracking efficiency ($96\%$ and uniform in $s_\pi$) was evaluated using  \pic$\pi^0$ and \pic
events 
identified by calorimeter information  plus
the presence of one fitted track.
The single-track efficiency as a function of $p_\pi$ and  $\theta_\pi$ was
 compared with Monte Carlo simulation; the difference, on the order of $0.3\%$ and 
flat in momentum, was taken as the systematic error of this procedure.

\item \textit {Vertex efficiency}:
The efficiency of the vertex finding algorithm has been evaluated
via Monte Carlo simulation and checked with a sample of \pic$\pi^0$ and \pic\gam\ events 
obtained from data.
The absolute vertex efficiency at low energies is $91\%$ and is increasing
up to $97\%$ at high values of $s_\pi$.
An uncertainty of 0.3\%, uniform in $s_\pi$, is taken as the contribution 
to the systematic error for this efficiency. 
%

\item \textit {Pion identification}:
The efficiency for $\pi/e$ separation has been evaluated by selecting $\ppg$ 
~events on the basis of one track and examining the distribution of the 
likelihood estimator for the other one. 
In the analysis, only one track is required to satisfy 
the likelihood requirement, for which the efficiency is $> 99.9\%$. Therefore, no correction for the inefficiency on pion identification 
needs to be applied; the contribution to the systematic error 
is taken to be $~0.1\%$. 

\item \textit {Track mass}:
The efficiency of the $\mtrk$ requirement is obtained as  a by-product
of the background evaluation; the
result of the fit provides the efficiency in each $s_\pi$ bin.
However, this efficiency 
 depends upon the treatment of 
multi-photon processes in the Monte Carlo simulation. 
The $\mtrk$ efficiency 
 has been obtained with our reference Monte Carlo simulation,
which uses PHOKHARA. To check the efficiency
determination we have compared
PHOKHARA with
  BABAYAGA~\cite{babayaga}, which is the generator used 
for the luminosity measurement. In the latter generator, ISR is
treated using the parton-shower approach. The resulting value for the
$\mtrk$ efficiency differs from that evaluated with
PHOKHARA by $0.2\%$. 
Effects on  the efficiency from the 
simultaneous emission of an ISR and a FSR photon are discussed 
in  Sec.~\ref{FSR}. 
\end{itemize}

\subsection{Unfolding of the mass resolution}
\label{Sec:Unfold}
\begin{figure}[ht]
\begin{center}
\mbox{\epsfig{file=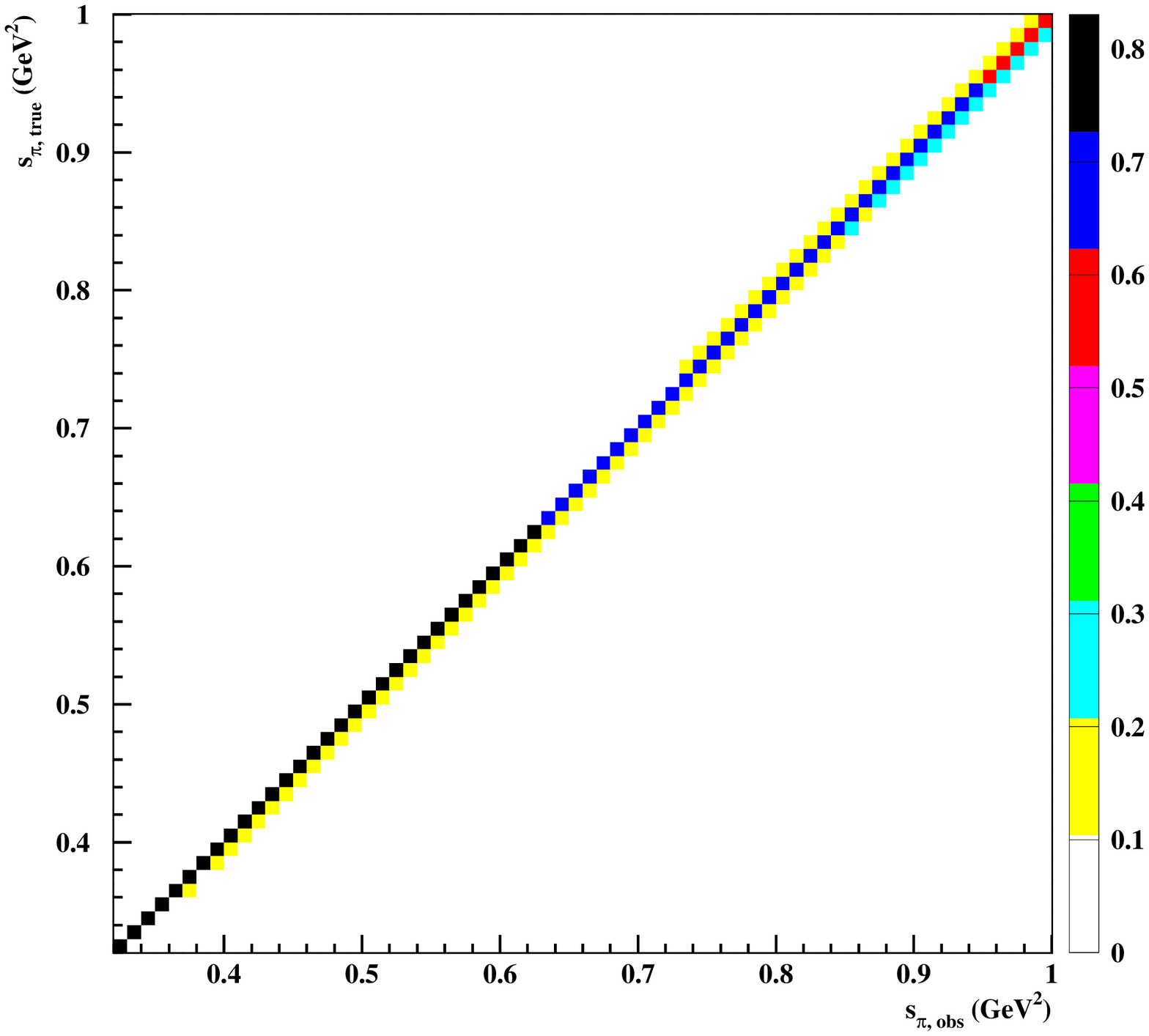,width=6.5cm}}
\mbox{\epsfig{file=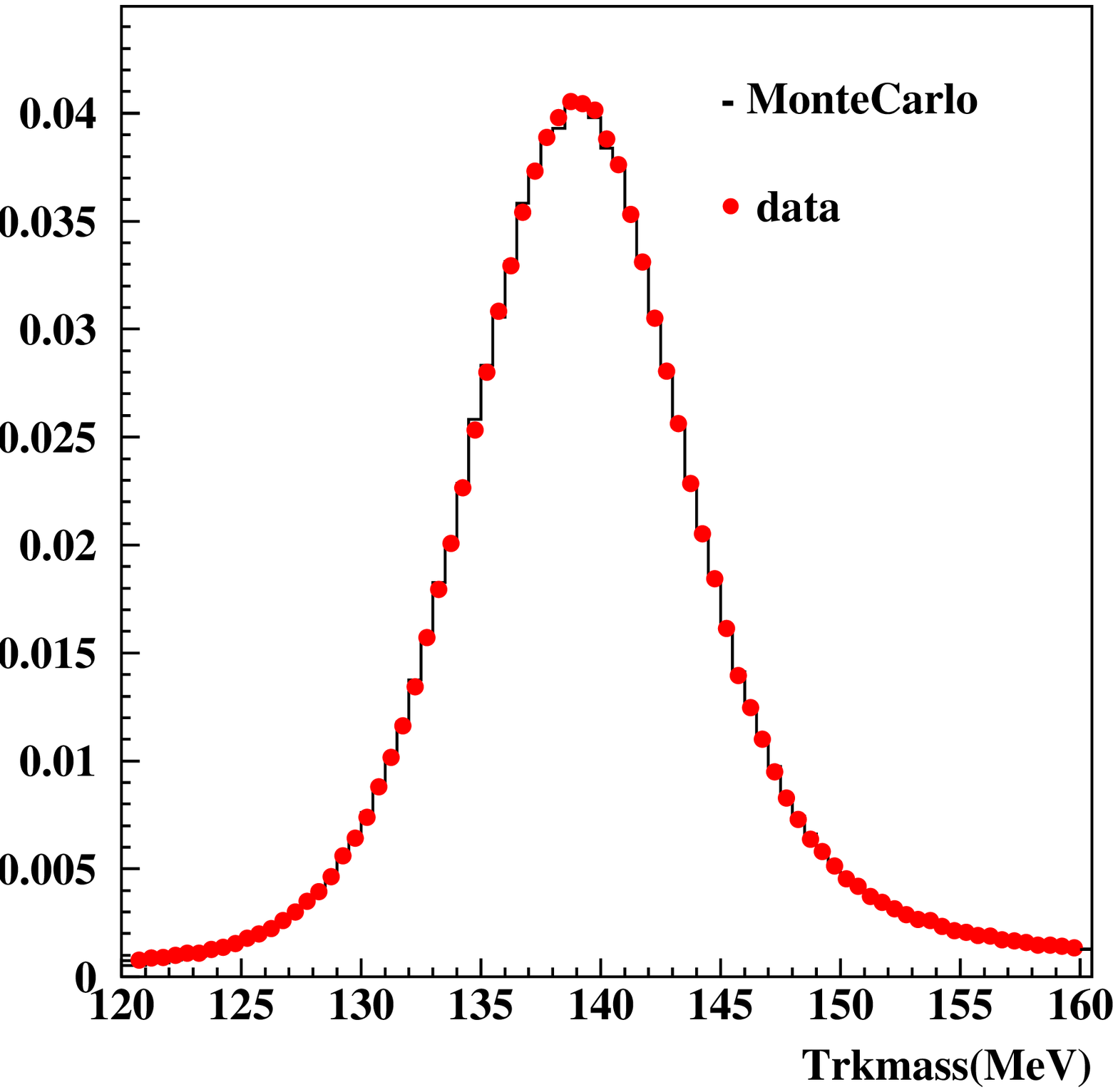,width=6.5cm}}
\caption{Left: Smearing matrix representing the correlation
between generated ($s_{\pi,{\rm true}}$) and reconstructed ($s_{\pi,{\rm obs}}$) values for $s_\pi$; the high precision
of the DC results in an almost diagonal matrix. Right: 
track mass distribution expected from the Monte Carlo simulation compared to the experimental one.}
\label{fig:matrix}
\end{center}
\end{figure}
\def\minus{\hbox{$-$}}
To obtain $d \sigma_{\pi \pi \gamma}/d s_\pi$ 
as a function of the true value of $s_\pi$, we unfold the mass 
resolution from the measured $s_\pi$ distribution.
The measured value of $s_{\pi,{\rm obs}}$ is related to the true 
value via the  resolution matrix 
{\bf G}($s_{\pi,{\rm true}}-s_{\pi,{\rm obs}}|s_{\pi,{\rm true}}$),
which   has been obtained by a Monte Carlo simulation 
carefully tuned to reproduce the data. 
The resolution matrix is nearly diagonal, as can be seen in 
in Fig.~\ref{fig:matrix}, left.
A comparison of the track-mass distribution for data and Monte Carlo 
events is shown Fig.~\ref{fig:matrix}, right. 



Unfolding of the spectrum is performed using GURU~\cite{hoecker}, 
an unfolding program based on the singular value decomposition (SVD).
We found that the systematic error due to  unfolding 
is dominated by the uncertainty on  the value chosen to regularize
 the procedure itself. Table~\ref{tab:syseff2} shows 
the systematic uncertainty as function of $s_{\pi}$, introduced into the 
$\ppg$ spectrum due to the unfolding.
These values are taken to be 
correlated errors, and translate into a 0.2\% systematic 
uncertainty
on $a_\mu$.\footnote{This value should
 be considered as an overestimate of the real effect introduced 
by the unfolding procedure on $a_\mu$, as discussed in~\cite{kn}.}

In addition, the unfolding procedure correlates the statistical errors in the
$\ppg$ spectrum (see also Sec.~\ref{lab:res}).

\subsection{Acceptance correction}
After all corrections discussed above, we obtain the spectrum for
$\ppg$ events defined by the acceptance requirements  
$50^\circ<\theta_{\pi}<130^\circ$,
$\theta_{\pi\pi}<15^\circ$ or $\theta_{\pi\pi}>165^\circ$, $p_T>160$ MeV,
and $p_z>90$ MeV. To derive the cross section 
for the process $e^+ e^- \rightarrow \pi^+ \pi^- \gamma$ with
$\theta_{\pi\pi}<15^\circ$ or $\theta_{\pi\pi}>165^\circ$, the
effects of the other requirements on the momentum and polar angle of the pions 
have been evaluated using PHOKHARA. The systematic error of $0.3\%$ on the acceptance fraction 
has been estimated by a comparison of data and Monte Carlo distributions.

\subsection{Luminosity measurement}
The integrated luminosity is measured with the KLOE detector itself
using very-large-angle Bhabha (VLAB) events.  The effective Bhabha
cross section at large angles ($55^o < \theta_{+,-} < 125^o$) is about
$430$ nb.  This cross section is large enough so that the statistical
error on the luminosity measurement is negligible.  The number of VLAB
candidates, $N_{\rm VLAB}$, is counted and normalized to the effective
Bhabha cross section, $\sigma_{\rm VLAB}^{\rm MC}$, obtained by Monte Carlo simulation, after
subtraction of the background, $\delta_{\rm Bkg}$:

\begin{equation}
\int{\mathcal L}dt =
\frac{N_{\rm VLAB}(\theta_i)}{\sigma_{\rm VLAB}^{\rm MC}(\theta_i)}
 (1-\delta_{\rm Bkg}).
\end{equation}

The precision of the luminosity measurement depends on
the correct inclusion of higher-order terms in computing
the Bhabha cross section. We use the Bhabha 
event generator BABAYAGA~\cite{babayaga}, which has been
 developed explicitly for DA$\Phi$NE. In BABAYAGA, QED radiative
corrections are taken into account in the framework of the 
parton-shower method. 
The precision quoted is $0.5\%$.  
The result for the effective Bhabha cross section has been compared with
that from BHAGENF~\cite{berends,drago}, a full order-$\alpha$ event
generator. We find agreement to better than $0.2\%$. 

\begin{figure}[hpt!]
\begin{center}
\mbox{\epsfig{file=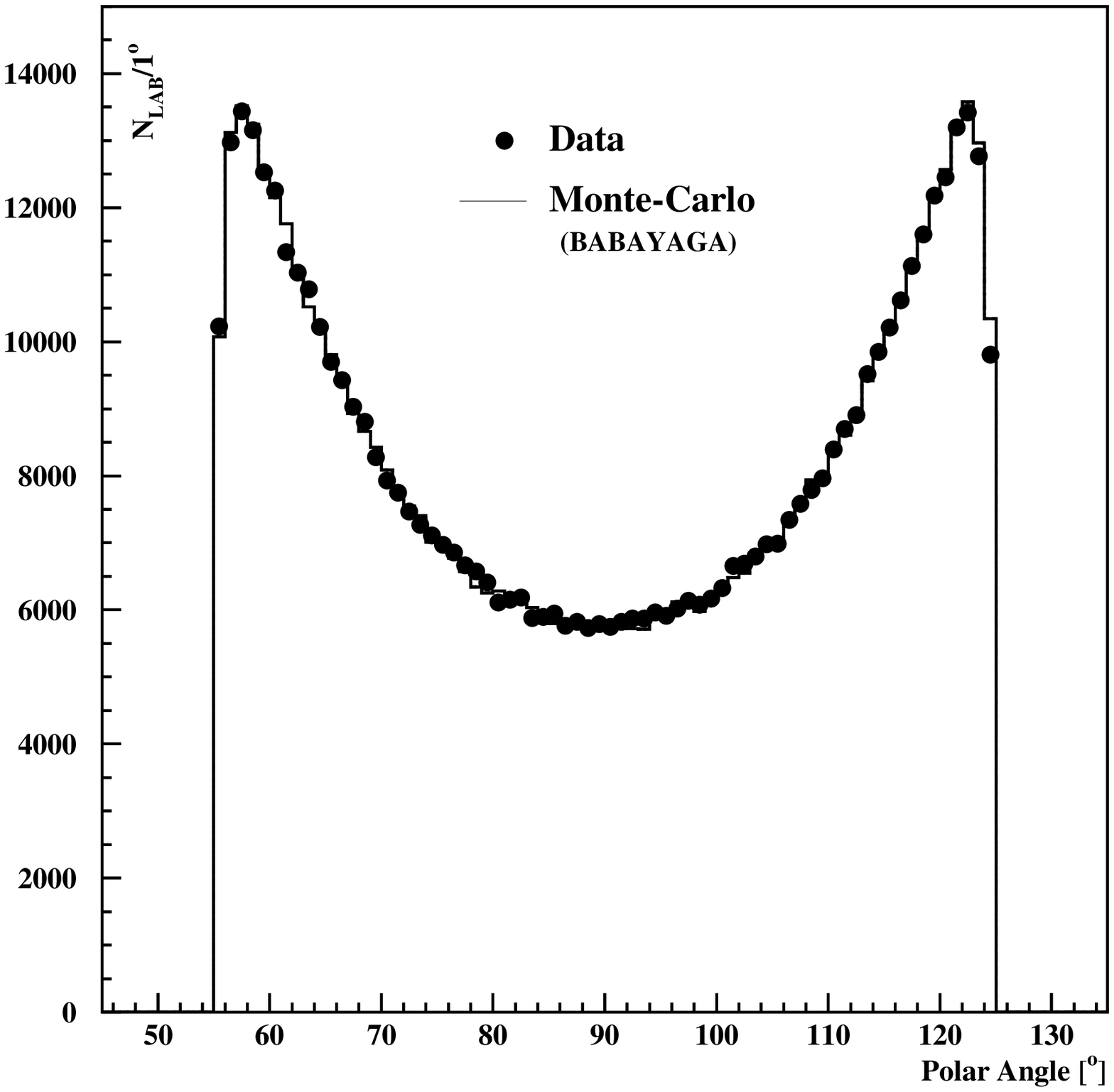,width=6.5cm}}
\mbox{\epsfig{file=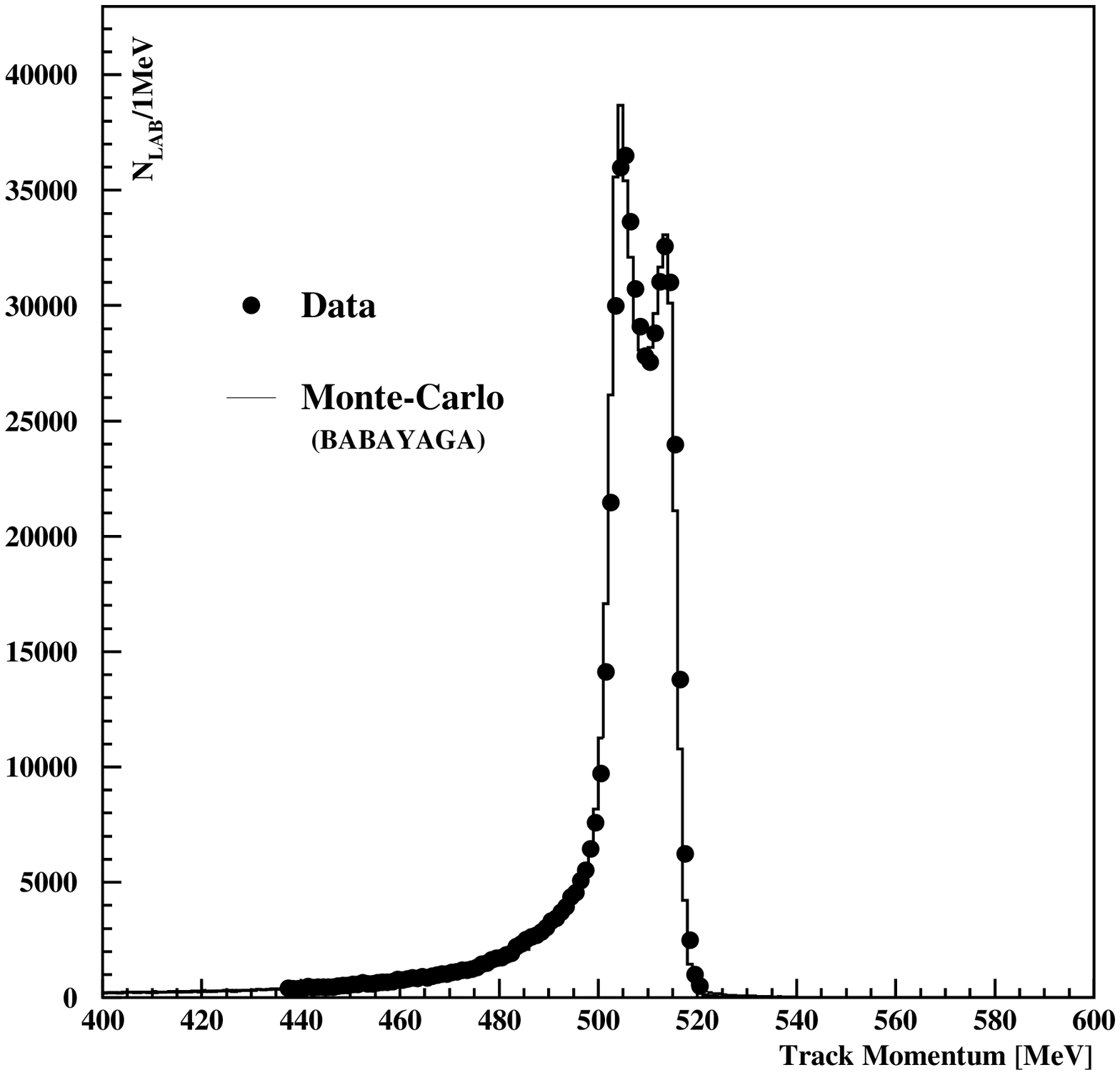,width=6.5cm}}
\caption{Data-Monte Carlo comparison
of the  $\theta_{+,-}$ (left) and $E_{+,-}$ (right) distributions
for Bhabha events selected at large angle as described
in the text.}
\label{fig:bhabha}
\end{center}
\end{figure}

VLAB events are selected with requirements on variables that are well reproduced by the KLOE Monte Carlo simulation.
The electron and positron polar angle reqiurements, 55\deg$\,<\theta_{+,-}<\,$125\deg, 
are based on the calorimeter clusters, while the energy requirements, $E_{+,-}>400$ MeV, are based on drift chamber information. 
The background from $\mu^+ \mu^- (\gamma)$, $\pi^+ \pi^- (\gamma)$ and $\pi^+ \pi^- \pi^0$ events is well below $1\%$ and is subtracted. 
All selection efficiencies (trigger, EmC cluster, DC tracking) 
are $>99\%$ as obtained by Monte Carlo simulation and confirmed with data. 
We obtain excellent agreement between the experimental distributions ($\theta_{+,-}$, $E_{+,-}$) and those obtained from Monte Carlo simulation, as
seen in Fig.~\ref{fig:bhabha}.
Finally, corrections are applied on a run-by-run basis for fluctuations in the center-of-mass energy of the machine and in the detector calibrations.
The experimental uncertainty in the acceptance due to all these effects is 0.3\%. We assign a total systematic error on the luminosity of 
$\delta{\mathcal L}=0.5\%_{\rm th}\oplus0.3\%_{\rm exp}$.
The luminosity measurement is independently checked  using \epm\to\gam\gam\ events. We find  agreement to within $0.2\%$. 

\subsection{$\ppg$ cross section}
Our results for the differential cross section 
 $d\sigma(e^+ e^- \rightarrow \pi^+ \pi^- \gamma)/ds_\pi$ with $0^\circ<\theta_\pi<180^\circ$
and $\theta_{\pi\pi}<15^\circ$, $\theta_{\pi\pi}>165^\circ$ are
plotted in 
Fig.~\ref{fig:sigppg}, left, and are presented in numerical form in the 
second  column of  Table~\ref{tab:results}.

\section{Extraction of 
 $\sigma(e^+ e^- \rightarrow \pi^+ \pi^-)$  and $|F_\pi(s)|^2$}
\begin{figure}[ht]
\begin{center}
\mbox{\epsfig{file=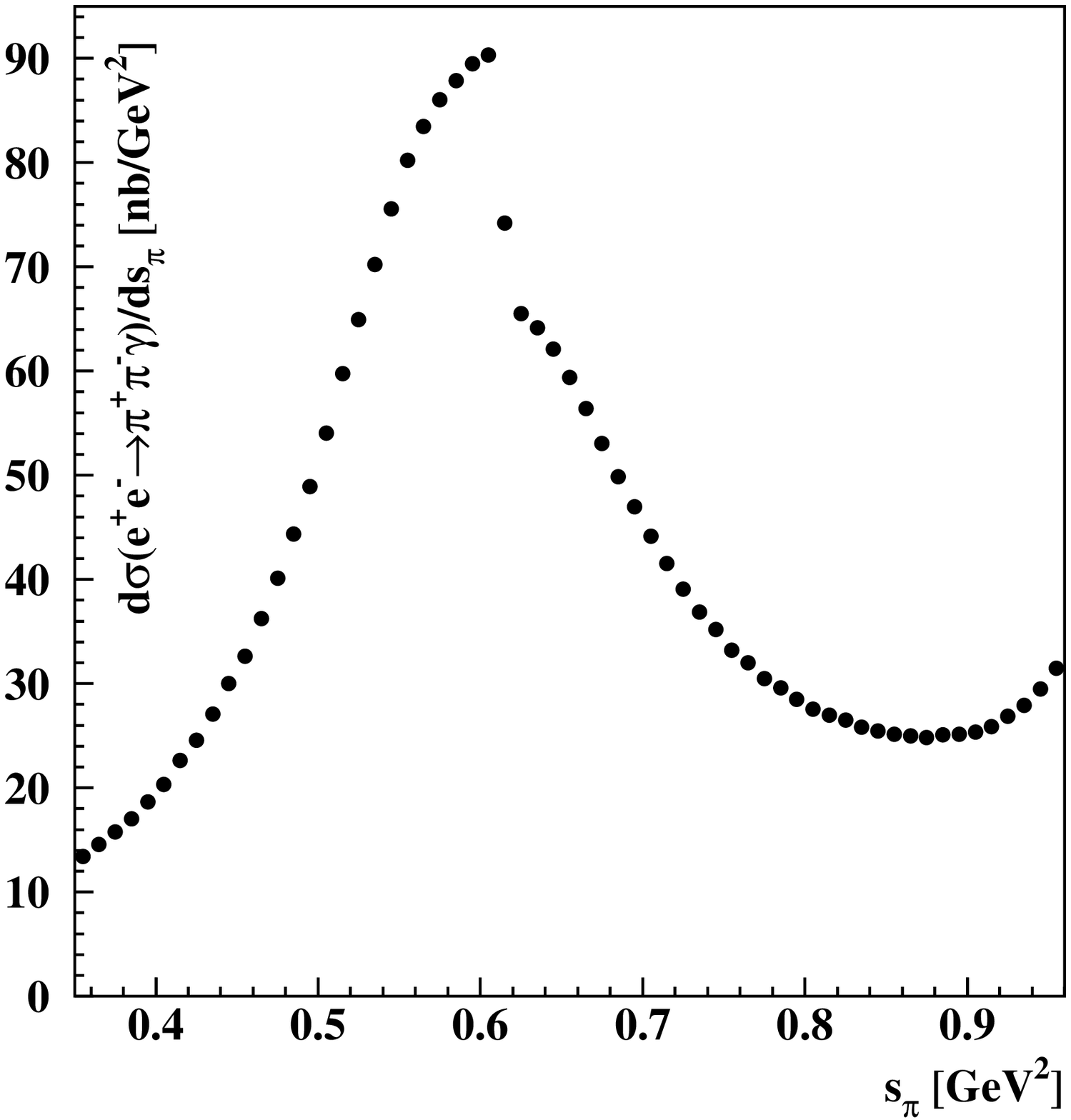,width=6.5cm}}
\mbox{\epsfig{file=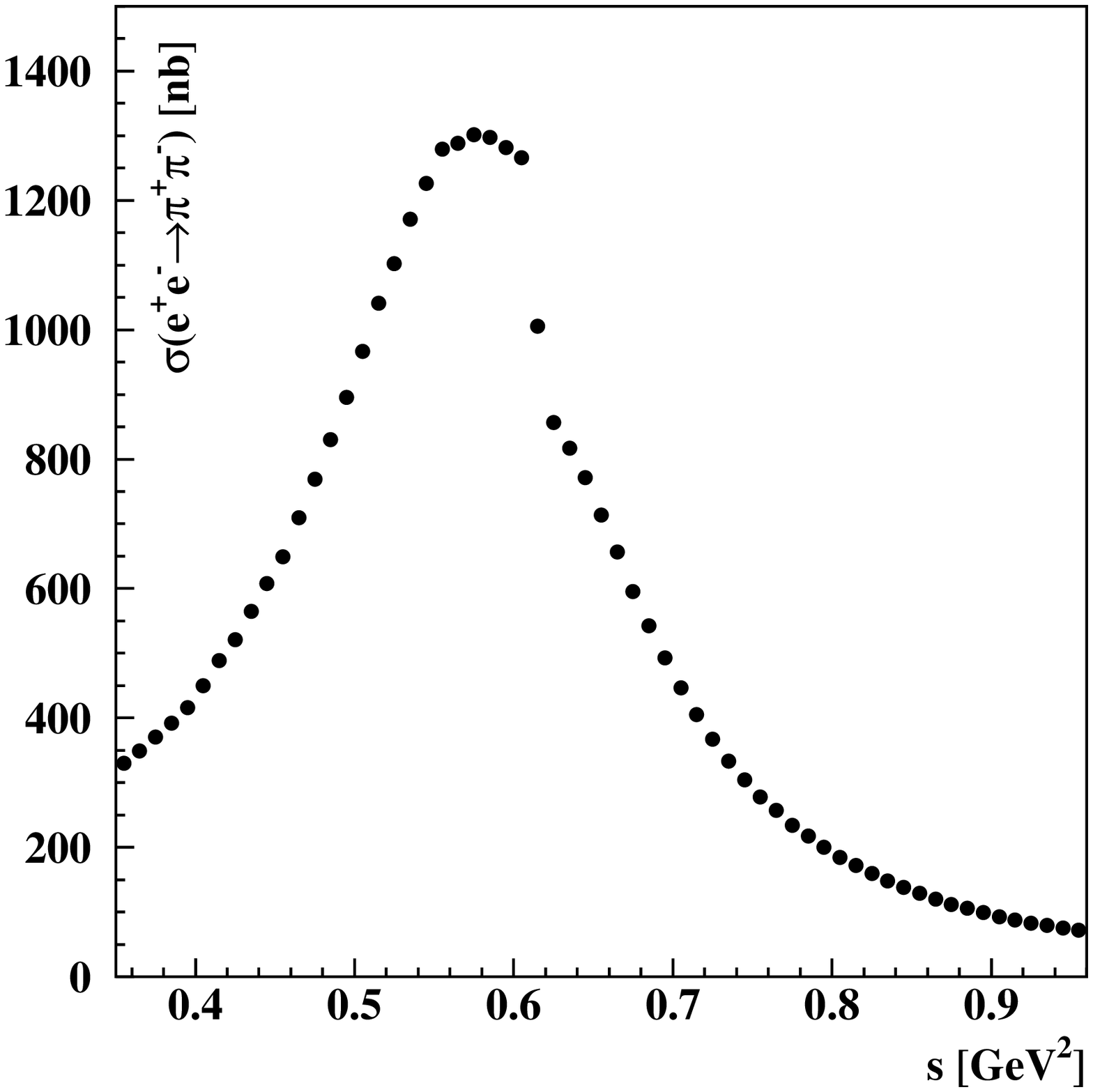,width=6.5cm}}
\caption{Left: differential cross section for the $e^+e^-\rightarrow \pi^+\pi^-
\gamma$ process, inclusive in $\theta_\pi$ and with
$\theta_{\pi\pi}<15^\circ$ ($\theta_{\pi\pi}>165^\circ$). 
Right:  cross section for $e^+e^-\rightarrow \pi^+\pi^-$.}
\label{fig:sigppg}
\end{center}
\end{figure}

In order to extract the $e^+ e^- \rightarrow \pi^+ \pi^- $ cross
section, the radiation function $H$ 
is needed (see Eq.~[\ref{eq:H}]).
This function is obtained from PHOKHARA, 
setting $F_\pi(s)=1$ and {\it switching
off} the vacuum polarization of the intermediate photon
in the generator. 
Applying Eq.~(\ref{eq:H}), 
and taking the FSR contribution into  account,
as described in the following section,
 the hadronic cross section as a function of the invariant mass of the
virtual photon, $s=m^2_{\gamma^*}$, is obtained, as
shown in
Fig.\ref{fig:sigppg}, right.

\subsection{FSR corrections}
\label{FSR}
Events with one or more photons emitted by the pions (FSR)
without any photons in the initial state 
must be considered as a 
background to our measurement.
Our event selection strongly suppresses the contribution of such events to
 well below 1\% over the entire range of $s_\pi$.

However,
events with the simultaneous emission of one photon from 
 the initial state and one photon from the final state  
 must be included 
in our selection in order for the $e^+e^-\rightarrow \pi^+\pi^-$ 
cross section to be 
inclusive with respect to FSR (see Ref.~\cite{Davier:2003pw} for details). 
More specifically, 
since the radiator function $H$ only describes the 
ISR part of the radiative corrections, 
the process $e^+e^-\to{\gamma^*} \to
\pi^+\pi^-\gamma_{\rm ISR} (\gamma_{\rm FSR})$,
with one photon from initial state and possibly another
from  the final state,  corresponds to 
$e^+e^-\to{\gamma^*}\to\pi^+\pi^- (\gamma_{\rm FSR})$ after the division by $H$.

Therefore,

\begin{equation}
\sigma(\epm\to{\rm \pi^{+}\pi^{-}(\gamma_{\rm FSR})})= 
\frac{\pi \alpha^2}{3 s} \beta_{\pi}^3  
\frac{d\sigma^{\pi\pi\gamma(\gamma_{\rm FSR})}}
{c_{s_\pi,s} A({s})d\sigma^{\pi\pi\gamma}({F_\pi=1})},
\label{eq:pion}
\end{equation}

where $d\sigma^{\pi\pi\gamma}({F_\pi=1})$ is the
 NLO cross section for $e^{+}e^{-} \rightarrow \ppg$ (initial state radiation only), inclusive in $\theta_{\pi\pi}$ and  $\theta_{\pi}$ 
under the assumption of pointlike pions, and corresponds to the quantity 
$H$ of Eq.~(\ref{eq:H}); $A({s})$ is the fraction of 
$\pi^+\pi^-\gamma_{\rm ISR} (\gamma_{\rm FSR})$ events  selected by the angular cuts 
$\theta_{\pi\pi}<15^\circ$ or $\theta_{\pi\pi}>165^\circ$, 
$50^\circ<\theta_{\pi}<130^\circ$ 
as a function of the  invariant mass $s$ of the virtual photon;
 and $c_{s_\pi,s}$ is a correction which must be applied 
due to the fact that, in the presence of simultaneous
 emission of initial-  and final-state  photons, 
 $s_\pi$ is not identical to 
$s$, as it is in the case of ISR only.
Both $A({s})$ and $c_{s_\pi,s}$
have been  obtained using the PHOKHARA Monte Carlo 
generator~\cite{fsrczyz}, which
simulates the  simultaneous emission of initial- and final-state  photons.


Note that  $\sigma(e^+e^-\rightarrow \pi^+\pi^-)$ 
is  obtained under the assumptions of ($i$) radiation 
emission from pointlike pions 
(the scalar QED model for FSR) and ($ii$) factorization, 
{\it i.e.}, the absence of interference effects between the initial and final 
states~\cite{fsrczyz}. 
We have used an alternative method which provides some
test of  the validity of the factorization 
ansatz and  a  valuable cross-check of the entire analysis.
In this method, we correct the observed $\pi\pi\gamma$ cross section
 for the relative amount
of FSR expected from PHOKHARA, obtaining, in this way, 
 a cross section that corresponds 
to ISR emission only. 
Next, we perform the event analysis, in which
the acceptance correction and track-mass efficiency
are taken from a Monte Carlo sample
 in which only ISR events are simulated.
After dividing by the radiator function $H$, the full 
({\it i.e.},  real and virtual) FSR corrections to the cross section 
$e^+e^-\to\pi^+\pi^-$ are 
applied~\cite{schwinger,hoefer}. 

The results for
$\sigma(e^+e^-\rightarrow \pi^+\pi^-)$ obtained with the
 two methods agree to within $\approx 0.2\%$.
Taking into account the additional uncertainty
arising  from the assumption of radiation from 
pointlike pions, we assign  
an error of  $0.3\%$ due to the FSR corrections 
discussed in this section.

\subsection{Vacuum polarization corrections}

To obtain the pion form factor and the {\it bare}
cross section, 
leptonic and hadronic vacuum
polarization contributions in the photon propagator must be subtracted.
This can be done by correcting the cross section for the running of
$\alpha$ as follows:
\begin{equation}
\sigma_{\rm bare} = \sigma_{\rm dressed} \left( \frac{\alpha(0)}{\alpha(s)}
\right)^2.
\end{equation}
While the leptonic contribution $\Delta\alpha_{\rm lep}(s)$ 
can be analytically calculated, 
for the hadronic contribution, $\Delta\alpha_{\rm had}(s)$, 
we have used  $\sigma_{\rm had}(s)$ values measured previously~\cite{vacpol}.

\begin{figure}[h!]
\begin{center}
\mbox{\epsfig{file=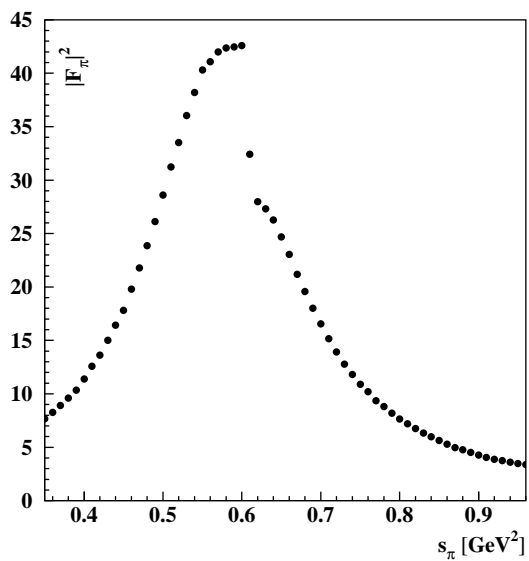,width=8cm}}
\caption{Pion form factor.}
\label{fig:fpi}
\end{center}
\end{figure}

The pion form factor $|F_\pi(s)|^2$ obtained after additional subtraction of FSR
is shown in Fig.~\ref{fig:fpi}. Note that in this case, since the FSR effects have been removed, $s_\pi=s$.


\section{Results}
\label{lab:res}
Our results are summarized in Table~\ref{tab:results},
which lists: 
\begin{itemize} 
\item the differential cross section $d\sigma(e^+ e^- \rightarrow \pi^+ \pi^-\gamma)/ds_\pi$ 
as a function of the invariant mass
 of the di-pion system, $s_\pi$,
  in the angular region $\theta_{\pi\pi} < 15^\circ$ or
  $\theta_{\pi\pi} >165^\circ$, $0^\circ < \theta_\pi < 180^\circ$;
\item the physical cross section  $\sigma(e^+ e^- \rightarrow \pi^+ \pi^-)$,
 which includes
  FSR and vacuum polarization effects, as a function of the invariant mass of 
the virtual photon $s$;
\item the pion form factor with FSR and vacuum polarization effects removed,
 as a function of $s$ (equal to $s_\pi$ in this case).
\end{itemize}

\noindent The errors given in Table~\ref{tab:results} 
are statistical only, while the common systematic error
is shown in Tables~\ref{tab:syseff0},~\ref{tab:syseff2}, 
and~\ref{tab:syseff3}. It should be noted that the 
statistical errors account only for the diagonal elements of 
the covariance matrix. The bin-by-bin errors are correlated as a result of the unfolding procedure; for error propagation, as 
for example in the  calculation of $a_{\mu}$ (see below), 
the  covariance matrix must be used.

The unfolding procedure is necessary in order to provide 
a table of data points at meaningful values of $s_\pi$.
However, the procedure itself introduces additional systematic 
uncertainties because of the numerical instability of the problem.
For the comparison of our data with a specific theoretical
 prediction, we strongly recomend fitting our observed spectrum 
with a convolution of the theoretical curve 
and the detector response matrix, which is available upon request~\cite{matr}.

\def\plm{\ifm{\pm}}  \def\km{\kern-1.5mm}  \def\kak{\km&\km}
\begin{table}[bp]
\centering
\renewcommand{\arraystretch}{1.15}
\begin{tabular}{||c|c|c|c||c|c|c|c||}
\hline
\km$s_\pi$\km&\pic\gam&\pic&\parbox{1cm}{\vglue3mm$|F_\pi(s)|^2$\vglue-3mm}&%
\km$s_\pi$\km&\pic\gam&\pic&\parbox{1cm}{\vglue3mm$|F_\pi(s)|^2$\vglue-3mm}\\
\km GeV$^2$\km& nb/GeV$^2$    &   nb            &    &
\km GeV$^2$\km& nb/GeV$^2$    &   nb            &    \\
 \hline
\km0.35\kak13.40\plm0.24\kak330\plm7\kak7.68\plm0.16\km&%
\km0.65\kak59.40\plm0.28\kak714\plm4\kak24.69\plm0.15\km\\
\km0.36\kak14.59\plm0.24\kak349\plm7\kak8.26\plm0.16\km&%
\km0.66\kak56.38\plm0.24\kak657\plm4\kak23.05\plm0.14\km\\
\km0.37\kak15.78\plm0.24\kak370\plm7\kak8.92\plm0.16\km&%
\km0.67\kak53.04\plm0.23\kak595\plm4\kak21.18\plm0.13\km\\
\km0.38\kak17.04\plm0.24\kak392\plm6\kak9.60\plm0.16\km&%
\km0.68\kak49.87\plm0.26\kak543\plm4\kak19.57\plm0.13\km\\
\km0.39\kak18.63\plm0.23\kak416\plm6\kak10.35\plm0.15\km&%
\km0.69\kak46.98\plm0.22\kak493.2\plm3.1\kak18.02\plm0.11\km\\
\km0.40\kak20.34\plm0.27\kak450\plm7\kak11.40\plm0.17\km&%
\km0.70\kak44.16\plm0.21\kak447.0\plm2.9\kak16.54\plm0.11\km\\
\km0.41\kak22.64\plm0.24\kak489\plm6\kak12.59\plm0.16\km&%
\km0.71\kak41.54\plm0.19\kak405.0\plm2.6\kak15.17\plm0.10\km\\
\km0.42\kak24.56\plm0.27\kak521\plm7\kak13.63\plm0.18\km&%
\km0.72\kak39.05\plm0.21\kak367.1\plm2.6\kak13.92\plm0.10\km\\
\km0.43\kak27.07\plm0.28\kak564\plm7\kak15.01\plm0.18\km&%
\km0.73\kak36.87\plm0.17\kak333.3\plm2.2\kak12.78\plm0.08\km\\
\km0.44\kak29.99\plm0.27\kak608\plm7\kak16.43\plm0.18\km&%
\km0.74\kak35.20\plm0.18\kak304.6\plm2.0\kak11.81\plm0.08\km\\
\km0.45\kak32.65\plm0.28\kak649\plm7\kak17.82\plm0.19\km&%
\km0.75\kak33.22\plm0.16\kak277.8\plm1.8\kak10.89\plm0.07\km\\
\km0.46\kak36.24\plm0.27\kak710\plm7\kak19.79\plm0.18\km&%
\km0.76\kak31.99\plm0.16\kak257.2\plm1.7\kak10.19\plm0.07\km\\
\km0.47\kak40.10\plm0.29\kak769\plm7\kak21.78\plm0.20\km&%
\km0.77\kak30.51\plm0.17\kak233.8\plm1.7\kak9.37\plm0.07\km\\
\km0.48\kak44.34\plm0.31\kak830\plm7\kak23.86\plm0.20\km&%
\km0.78\kak29.60\plm0.16\kak217.7\plm1.6\kak8.82\plm0.06\km\\
\km0.49\kak48.94\plm0.28\kak895\plm7\kak26.11\plm0.20\km&%
\km0.79\kak28.52\plm0.13\kak200.3\plm1.3\kak8.20\plm0.05\km\\
\km0.50\kak54.1\plm0.4\kak967\plm8\kak28.60\plm0.23\km&%
\km0.80\kak27.53\plm0.14\kak184.5\plm1.3\kak7.63\plm0.05\km\\
\km0.51\kak59.77\plm0.32\kak1041\plm7\kak31.23\plm0.22\km&%
\km0.81\kak27.00\plm0.14\kak172.1\plm1.2\kak7.20\plm0.05\km\\
\km0.52\kak64.93\plm0.32\kak1102\plm7\kak33.50\plm0.22\km&%
\km0.82\kak26.48\plm0.13\kak160.0\plm1.1\kak6.76\plm0.05\km\\
\km0.53\kak70.24\plm0.35\kak1171\plm8\kak36.05\plm0.23\km&%
\km0.83\kak25.84\plm0.15\kak148.4\plm1.1\kak6.33\plm0.05\km\\
\km0.54\kak75.6\plm0.4\kak1226\plm8\kak38.20\plm0.26\km&%
\km0.84\kak25.45\plm0.13\kak138.5\plm1.0\kak5.97\plm0.04\km\\
\km0.55\kak80.2\plm0.4\kak1279\plm8\kak40.32\plm0.24\km&%
\km0.85\kak25.16\plm0.13\kak129.2\plm0.9\kak5.63\plm0.04\km\\
\km0.56\kak83.47\plm0.35\kak1288\plm7\kak41.07\plm0.24\km&%
\km0.86\kak24.96\plm0.12\kak120.3\plm0.8\kak5.29\plm0.04\km\\
\km0.57\kak86.06\plm0.34\kak1302\plm7\kak41.98\plm0.23\km&%
\km0.87\kak24.81\plm0.15\kak111.8\plm0.9\kak4.97\plm0.04\km\\
\km0.58\kak87.85\plm0.34\kak1297\plm7\kak42.36\plm0.23\km&%
\km0.88\kak25.09\plm0.14\kak106.3\plm0.8\kak4.774\plm0.035\km\\
\km0.59\kak89.5\plm0.4\kak1282\plm7\kak42.46\plm0.24\km&%
\km0.89\kak25.17\plm0.12\kak99.5\plm0.7\kak4.516\plm0.030\km\\
\km0.60\kak90.31\plm0.35\kak1266\plm7\kak42.58\plm0.23\km&%
\km0.90\kak25.37\plm0.13\kak93.1\plm0.6\kak4.269\plm0.030\km\\
\km0.61\kak74.20\plm0.35\kak1006\plm6\kak32.43\plm0.20\km&%
\km0.91\kak25.86\plm0.12\kak87.6\plm0.6\kak4.059\plm0.027\km\\
\km0.62\kak65.49\plm0.28\kak857\plm5\kak27.99\plm0.16\km&%
\km0.92\kak26.87\plm0.14\kak83.0\plm0.6\kak3.886\plm0.026\km\\
\km0.63\kak64.14\plm0.28\kak817\plm5\kak27.32\plm0.16\km&%
\km0.93\kak27.94\plm0.14\kak79.1\plm0.5\kak3.741\plm0.025\km\\
\km0.64\kak62.09\plm0.26\kak772\plm4\kak26.27\plm0.15\km&%
\km0.94\kak29.49\plm0.16\kak75.3\plm0.5\kak3.599\plm0.025\km\\
\hline
\end{tabular}
\caption{Cross sections $d\sigma(e^+ e^- \rightarrow \pi^+ \pi^-\gamma)/ds_\pi$, $\sigma(e^+ e^- \rightarrow \pi^+ \pi^-)$ and the 
pion form factor 
in 0.01 GeV$^2$ intervals where the value given indicates the lower bound.
Note that while the $\ppg$ cross section is given as a function of $s_\pi$,
the $\pi\pi$ cross section and $|F_\pi|^2$ are given as functions of the invariant mass $s$ of the intermediate photon $\gamma^*$.}
\label{tab:results}
\end{table}

\begin{table}
\begin{center}
\begin{tabular}{||c|c|c|c|c|c|c|c|c|c|c||}
\hline
$s$ (GeV$^2$) &  0 &  1 &  2 &  3 &  4 &  5 &  6 &  7 &  8 &  9 \\
\hline
0.3\dd&     &     &     &     &     &0.8&0.7&0.6&0.6&0.5\\
0.4\dd&0.5&0.4&0.4&0.4&0.4&0.4&0.4&0.4&0.4&0.4\\
0.5\dd&0.3&0.3&0.3&0.3&0.3&0.3&0.3&0.3&0.3&0.3\\
0.6\dd&0.3&0.3&0.3&0.3&0.3&0.3&0.3&0.3&0.3&0.3\\
0.7\dd&0.3&0.2&0.3&0.3&0.3&0.3&0.3&0.2&0.2&0.2\\
0.8\dd&0.2&0.2&0.2&0.2&0.2&0.2&0.2&0.2&0.2&0.2\\
0.9\dd&0.2&0.2&0.2&0.2&0.2&     &     &     &     &     \\ 
\hline
\end{tabular}
\caption{Bin-by-bin correlated systematic error in \% due to background subtraction in 
 0.01 GeV$^2$ intervals. The
indicated values for $s$ represent the lower bin edge.}
\label{tab:syseff0}
\end{center}
\end{table}

\begin{table}
\begin{center}
\begin{tabular}{||c|c|c|c|c|c|c|c|c||}
\hline
$s$ (GeV$^2$) &  0.58 & 0.59 & 0.6 & 0.61 & 0.62 & 0.63 & 0.64 & 0.65 \\
\hline
$\delta_{unf}$ & 0.4 & 0.9 & 1.4 & 3.6 & 0.9 & 0.8 & 0.5 & 0.4 \\
\hline
\end{tabular}
\caption{Bin-by-bin correlated systematic error in \% on 
\mbox{$d\sigma (e^{+}e^{-}
\rightarrow \pi^{+}\pi^{-}\gamma)/ds_{\pi}$} and
\mbox{$\sigma(e^+e^-\rightarrow \pi^+\pi^-)$} due to unfolding in 0.01 GeV$^2$ intervals. The
indicated values for $s$ represent the lower bin edge.}
\label{tab:syseff2}
\end{center}
\end{table}

\begin{table}
\begin{center}
\begin{tabular}{||l|c|c|c||}
\hline
 & $\sigma_{\pi\pi\gamma}$ & $\sigma_{\pi\pi}$ & $|F_\pi|^2$ \\
\hline
\hline
Acceptance & \multicolumn{3}{|c||}{0.3 \% flat in $s_\pi$} \\
Trigger & \multicolumn{3}{|c||}{$\mbox{exp}(0.43-4.9s_\pi [{\rm GeV}^2])$ \% + 0.08 \%} \\
Reconstruction Filter & \multicolumn{3}{|c||}{0.6 \% flat in $s_\pi$} \\
Tracking & \multicolumn{3}{|c||}{0.3 \% flat in $s_\pi$} \\
Vertex & \multicolumn{3}{|c||}{0.3 \% flat in $s_\pi$} \\
Particle ID & \multicolumn{3}{|c||}{0.1 \% flat in $s_\pi$} \\
Trackmass & \multicolumn{3}{|c||}{0.2 \% flat in $s_\pi$} \\
\hline
\hline
Luminosity &  \multicolumn{3}{|c||}{0.6 \% flat in $s_\pi$} \\ \cline{2-4}
FSR resummation & - & \multicolumn{2}{|c||}{0.3 \% } \\ \cline{2-4}
Radiation function ($H(s_{\pi})$) & - & \multicolumn{2}{|c||}{0.5 \% } \\ \cline{2-4}
Vacuum Polarization & - & - & 0.2 \% \\ \cline{2-4} 
\hline
\end{tabular}
\caption{List of completely bin-by-bin correlated systematic effects.}
\label{tab:syseff3}
\end{center}
\end{table}

\begin{table}
\begin{center}
\begin{tabular}{||l|c||}
\hline
Acceptance & 0.3 \% \\
Trigger & 0.3 \% \\
Reconstruction Filter & 0.6 \% \\
Tracking & 0.3 \% \\
Vertex & 0.3 \% \\
Particle ID & 0.1 \% \\
Trackmass & 0.2 \% \\
Background subtraction & 0.3 \% \\
Unfolding & 0.2 \% \\
\hline
Total exp systematics & 0.9 \% \\
\hline
\hline
Luminosity & 0.6 \% \\
Vacuum Polarization &  0.2 \% \\
FSR resummation & 0.3 \% \\
Radiation function ($H(s_{\pi})$) & 0.5 \% \\
\hline
Total theory systematics & 0.9 \% \\
\hline
\end{tabular}
\caption{List of systematic errors on $a_{\mu}$.}
\label{tab:syseff1}
\end{center}
\end{table}

The $\sigma(e^+ e^- \rightarrow \pi^+ \pi^-)$ cross section, 
divided by the vacuum polarization,
 has been used to evaluate the contribution to $a_\mu^{\rm had}$
due to the \pic\ channel in the energy range $0.35<s_\pi<0.95$ \,GeV$^2$.
The resulting value (in 10$^{-10}$ units) is 
\begin{equation}
a_\mu^{\rm \pi\pi}(0.35,0.95)  = 
388.7 \pm 0.8_{\rm stat} \pm 3.5_{\rm syst} \pm 3.5_{\rm th}. 
\end{equation}
The various contributions to the systematic error on $a_\mu$ are listed in 
Table~\ref{tab:syseff1}.

\section{Conclusions} 

We have measured the cross section for
 the process \epm\to~\pic\gam\ 
with the  pion system emitted at small polar
angles with respect to the electron or positron beam ($\theta_{\pi\pi}<15^\circ$, $\theta_{\pi\pi}>165^\circ$)
in the energy region $0.35<s_\pi<0.95$ GeV$^2$.
Using Eq.~(\ref{eq:H}), we have derived the cross section for
the process \epm\to~\pic , as listed in Table~\ref{tab:results}.
These values, corrected for the vacuum polarization,
can be used to derive part of the hadronic contribution to the
muon anomalous magnetic moment with a negligible statistical
error and with a systematic error of $0.9\% {\rm (exp)} \oplus
0.9\%{\rm (th)}$. 

Future improvements are expected using data taken in 2002, where more
stable background conditions and an improved trigger logic should
allow for a considerable reduction of the systematic effects stemming
from the offline reconstruction filter and the trigger. A similar
 analysis, applied to events with $\theta_{\pi\pi}$ at
larger angles, can probe the energy region down to threshold.
Moreover, improved Monte Carlo generators
both for the luminosity measurement and for the ISR process are
expected to be available 
in the near future, which will help to reduce the theoretical
contribution to the systematic error.

\section*{Acknowledgements}
We would like to thank Carlo Michel Carloni Calame, Henryk Czy{\.z}, Axel H\"ofer, Stanislaw Jadach, 
Fred Jegerlehner, Johann K\"uhn, Guido Montagna, and Germ{\'a}n Rodrigo 
for numerous useful discussions.

We thank the DAFNE team for their efforts in maintaining 
low background running 
conditions and their collaboration during all data-taking. 
We want to thank our technical staff: 
G.F. Fortugno for his dedicated work to ensure an efficient operation of 
the KLOE Computing Center; 
M. Anelli for his continous support to the gas system and the safety of the
detector; 
A. Balla, M. Gatta, G. Corradi, and G. Papalino for the maintenance of the
electronics;
M.Santoni, G.Paoluzzi, and R.Rosellini for the general support to the
detector; 
C.Pinto (Bari), C.Pinto (Lecce), C.Piscitelli, and A.Rossi for
their help during major maintenance periods.
This work was supported in part by DOE grant DE-FG-02-97ER41027; 
by EURODAPHNE, contract FMRX-CT98-0169; 
by the German Federal Ministry of Education and Research (BMBF) contract 06-KA-957; 
by Graduiertenkolleg `H.E. Phys. and Part. Astrophys.' of Deutsche 
Forschungsgemeinschaft,
Contract No. GK 742 and 'Emmy-Noether Research group', Contract No. DE839/1; 
by INTAS, contracts 96-624, 99-37; 
and by TARI, contract HPRI-CT-1999-00088.


\bibliographystyle{elsart-num}
\bibliography{PPG_PLB}
\end{document}

\end{document}